\documentclass[twocolumn,showpacs,preprintnumbers,prd]{revtex4}
\usepackage{psfrag,graphicx,epsfig,amsmath,amssymb,bm,color}
\usepackage[bookmarks=false]{hyperref}

\def\bm#1{\mbox{\boldmath$#1$\unboldmath}}

\def\Mkk{M_{\rm KK}}
\def\etal {\em et al.}
\newcommand{\Q}{\mathcal Q}
\newcommand{\eps}{\epsilon}
\newcommand{\ord}{{\cal O}}

\newcommand{\sws}{s_w^2}

\newcommand{\cws}{c_w^2}

\newcommand{\beq}{\begin{equation}}
\newcommand{\eeq}{\end{equation}}
\newcommand{\nr}{\nonumber}

\begin{document}

\preprint{MZ-TH/11-05} 


\title{\boldmath
Randall-Sundrum Corrections to the Width Difference and CP-Violating Phase in $B^0_s$-Meson Decays
\unboldmath}

\author{Florian Goertz and Torsten Pfoh} 

\affiliation{
Institut f\"ur Physik (THEP), Johannes Gutenberg-Universit\"at,
D-55099 Mainz, Germany
}

\begin{abstract}

We study the impact of the Randall-Sundrum setup on the width difference $\Delta\Gamma_s$ and the CP-violating phase $\phi_s$ in the $\bar B_s^0$-$B_s^0$ system. Our calculations are performed in the general framework of an effective theory, based on operator product expansion. The result can thus be used for many new physics models. We find that the correction to the magnitude of the decay amplitude $\Gamma_{12}^s$ is below $4\%$ for a realistic choice of input parameters. The main modification in the $\Delta\Gamma_s/\beta_s$-plane is caused by a new CP-violating phase in the mixing amplitude, which allows for a better agreement with the experimental results of CDF and D\O~ from $B_s^0\rightarrow J/\psi\phi\,$ decays. The best-fit value of the CP asymmetry $S_{\psi\phi}$ can be reproduced, while simultaneously the theoretical prediction for the semileptonic CP asymmetry $A_{\rm SL}^s$ can enter the $1\sigma$ range.
 
\end{abstract}

\date{\today}


\maketitle

\section{Introduction}
\label{sec:intro}

Within the search for new physics (NP) in the decay of $B_s^0$-mesons, an important 
observable is the width difference $\Delta\Gamma_s\equiv\Gamma^s_L-\Gamma^s_H$ between the
light and the heavy meson state. According to the above definition, $\Delta\Gamma_s$ 
happens to be positive in the Standard Model (SM).
It can be computed from the dispersive and absorptive
part of the $\bar B_s^0$-$B_s^0$ mixing amplitude, $M_{12}^s$ and $\Gamma_{12}^s\,$. 
To leading order in $|\Gamma_{12}^s|/|M_{12}^s|$ one finds the simple relation 
\cite{Buras:1984pq,Dunietz:1987yt}
\beq\label{eq:DeltaG}
\displaystyle
\Delta\Gamma_s= -\,\frac{2\,\text{Re}(M_{12}^s\Gamma_{12}^{s*})}{|M_{12}^s|} 
=2\,|\Gamma_{12}^s|\,\cos \phi_s\,.
\eeq
We define the relative phase $\phi_s$ between the mixing and the 
decay amplitude according to the convention
\beq\label{eq:phase}
\displaystyle
\frac{M_{12}^s}{\Gamma_{12}^s}= -\,\frac{|M_{12}^s|}{|\Gamma_{12}^s|}\;e^{i\phi_s}\,,\qquad
\phi_s=\arg(-M_{12}^s\Gamma_{12}^{s\,*})\,,
\eeq
for which the SM value is positive and explicitly given by  
$\phi_s^{\rm SM}=(4.2\pm1.4)\cdot 10^{-3}$ \cite{Lenz:2006hd}.
The combined experimental  results of CDF and D\O~ differ from the SM prediction in the
$(\beta_s^{J/\psi\phi},\Delta\Gamma_s)$-plane by about $2\sigma$ \cite{pubn}, whereas the latest
CDF results disagree by $1\sigma$ only \cite{pubn2}. Here, 
$\beta_s^{J/\psi\phi}\in[-\pi/2,\pi/2]$ is the CP-violating phase 
in the interference of mixing and decay, obtained from the time-dependent angular analysis 
of flavor-tagged $B_s^0\rightarrow J/\psi\phi\,$ decays.
In the SM it is given by \cite{Lenz:2006hd,Ligeti:2006pm}
\beq\label{eq:betasSM}
\displaystyle
\beta_s^{J/\psi\phi}=-\arg\left(- \frac {\lambda_{t}^{bs}}{\lambda_{c}^{bs}}\right)
= 0.020\pm0.005\,,
\eeq
with $\lambda_{q}^{bs}=V_{qb}V_{qs}^*\,$. 
In the presence of NP, $\Delta \Gamma_s$ will be modified \cite{Grossman:1996era,Dunietz:2000cr}.
We adopt the notation of \cite{Dighe:2010nj} and extend the SM relations according to
\begin{align}
\begin{split}
M_{12}^s&=M_{12}^{s\,\rm{SM}}+M_{12}^{s\,\rm{NP}} = M_{12}^{s\,\rm{SM}} R_M\, e^{i\phi_M}\,,\\
\Gamma_{12}^s&=\Gamma_{12}^{s\,\rm{SM}}+\Gamma_{12}^{s\,\rm{NP}} = 
\Gamma_{12}^{s\,\rm{SM}} R_\Gamma\, e^{i\phi_\Gamma}\,.
\end{split}
\end{align}
From (\ref{eq:DeltaG}) it follows that 
\beq\label{eq:DGammaapprox}
\Delta\Gamma_s=2\,|\Gamma_{12}^{s\,\rm{SM}}|\,R_\Gamma\,\cos(\phi_s^{\rm SM}+\phi_M-\phi_\Gamma)\,,
\eeq
where $\Delta\Gamma_s^{\rm SM}=(0.087\pm 0.021)\, {\rm ps}^{-1}$ \cite{Lenz:2011ti}.
A further important observable is the semileptonic CP asymmetry $A_{\rm SL}^s={\rm Im}(\Gamma_{12}^s/M_{12}^s)$. 
Including NP corrections, we find
\beq
A_{\rm SL}^s=\,\frac{|\Gamma_{12}^{s\,{\rm SM}}|}{|M_{12}^{s\,{\rm SM}}|}\,
\frac{R_\Gamma}{R_M}\;\sin (\phi_s^{\rm SM}+\phi_M-\phi_\Gamma)\,.
\eeq

Within the SM, the leading contribution to the dispersive part of the $\bar B_s^0$-$B_s^0$ 
mixing amplitude appears at the one loop level. If NP involves flavor-changing 
neutral currents (FCNCs) at tree level, these give rise to sizable corrections to the mass 
difference $\Delta m_{B_s}\equiv M_H^s-M_L^s=2\,|M_{12}^s|\,$ \cite{Buras:1984pq}.
In the context of Randall-Sundrum (RS) scenarios \cite{Randall:1999ee}, the corrections to $M_{12}^s$ have been
calculated in \cite{Blanke:2008zb,Bauer:2009cf}. See also \cite{Agashe:2004ay,Agashe:2004cp} for
a first estimate.

On the other hand, the presence of tree-level FCNCs and right-handed 
charged-current interactions give rise to new decay diagrams.
However, the NP corrections to the absorptive part of the amplitude are suppressed by 
$m_W^2/\Lambda^2$ with respect to the SM contribution, where $\Lambda$ is the NP mass scale. 
Thus, they are neglected in many NP studies.
Recently, model-independent estimates on $A_{\rm SL}^s$ in the presence of heavy
gluons have been presented in \cite{Datta:2010yq}, taking into 
account modifications in $\Gamma_{12}^s$. 
NP contributions from electroweak (EW) penguin operators as well as right-handed charged 
currents have not been considered. We find that the former can compete with or even 
dominate contributions from QCD penguins within the minimal RS model 
\cite{Casagrande:2008hr,Bauer:2009cf}, where part of the latter tend to give the dominant 
contribution to $\Gamma_{12}^{s\,{\rm RS}}$ for the most natural choice of input parameters.

This article is organized as follows. 
In the next section we briefly summarize the main features of the RS model.
We distinguish between two variants, the minimal and the custodial RS model
with protection of the $Zb_L\bar b_L$ vertex, each with a brane-localized Higgs.
Then we calculate the leading contributions to $\Gamma_{12}^s$ in the presence of NP, 
where we restrict ourselves to operators which are expected to give the dominant 
corrections for the models at hand.
A numerical scan across RS contributions is presented in Section \ref{sec:num}. 
Here, we evaluate $M_{12}^s$ and $\Gamma_{12}^s$ for $10000$ appropriate 
random sets of input parameters. 
Important constraints arise from the $\bar B_s^0$-$B_s^0$ oscillation frequency, which corresponds 
to the mass difference $\Delta m_{B_s}$, and the observable $\epsilon_K$.
The results are presented in the $\Delta\Gamma_s/\beta_s$- 
as well as the $A_{\rm SL}^s/S_{\psi\phi}\,$-plane. We conclude 
in Section \ref{sec:conc}. In a series of appendices we collect analytic results 
for RS Wilson coefficients needed in our computations.

\section{Features of the RS model}\label{sec:RSimp}

The RS model is formulated on a five-dimensional (5D) Anti-de Sitter space.
The compactified fifth dimension is an $S^1/Z_2$ orbifold, labeled by a 
dimensionless coordinate $\phi\in [-\pi,\pi]\,$.
The usual 4D space-time is rescaled by a so-called warp factor, 
such that length scales depend on the position in the extra dimension. The whole (5D)
space-time is called the bulk. The RS metric is given by
\beq
ds^2=e^{-2 kr|\phi|}\eta_{\mu\nu}\,dx^\mu dx^\nu-r^2d\phi^2\,,
\eeq
with $\eta_{\mu\nu}=\,$diag$(1,-1,-1,-1)$. Here, $k$ and $r$ denote the curvature
and the radius of the fifth dimension, which are of the order of the (inverse) Planck scale. 
The $Z_2$-parity identifies points $(x^\mu,\phi)$ and $(x^\mu,-\phi)$ and thus gives 
rise to boundaries at $\phi=0$ and $\pi$, which are called Planck/ultraviolet (UV) and 
TeV/infrared (IR) brane respectively. The RS model solves the gauge hierarchy problem by 
suppressing mass scales on the IR-brane. Explicitly, one achieves
\beq
M_{\rm IR}\equiv e^{-L} M_{\rm Pl}\equiv \eps\, M_{\rm Pl}\approx M_{\rm W}\,
\eeq
for $L\,\equiv kr\pi\approx 37$ $(\eps=10^{-16})$. Thus, the strong hierarchy between the 
Planck and the weak scale, $M_{\rm Pl}$ and $M_{\rm W}$, is understood by gravitational 
red-shifting, if the Higgs field is localized on or near the IR-brane.  
An effective four dimensional description is usually obtained via Kaluza-Klein (KK) 
decomposition, which replaces each 5D field by an infinite tower of massive 4D fields, 
each of them supplied with a so-called profile depending on $\phi$. 
Even fields under $Z_2$ (which in addition obey Neumann boundary conditions (BCs) on both branes)
posses a massless zero mode, which can however receive a mass via coupling to the Higgs.
Those light modes are interpreted as the SM fields. The masses of the additional heavy KK modes are of the order 
of the scale $\Mkk=k\eps\approx$ few TeV, which is identified with the cut-off $\Lambda$ 
of the effective low-energy theory. For instance, the mass of the first KK gluon
is given by $m^{(1)}\approx 2.45\, \Mkk$. Take care of the fact that some authors define $\Mkk$ as the mass of the
first excitation. Explicit formulas for the fermion- and gauge-boson profiles have first been 
given in \cite{Grossman:1999ra,Gherghetta:2000qt} and \cite{Davoudiasl:1999tf,Pomarol:1999ad}, respectively.
The warp factor can be used to generate fermion-mass hierarchies \cite{Grossman:1999ra,Gherghetta:2000qt,Huber:2000ie}. 
This is achieved by localizing the fields  differently in the bulk by an appropriate choice 
of the doublet/singlet 5D mass parameters $M_{{Q_i}/{q_i}}$, which are often called bulk masses. 

The appearance of tree level FCNCs is caused by the modified interactions between gauge and 
matter fields, which now contain overlap integrals of the corresponding profiles. 
If the gauge field possesses a mass, the overlap is flavor (and KK mode) dependent, giving rise to 
FCNCs when changing from the weak interaction to the mass eigenbasis. A crucial observation is that 
these non-universal overlap integrals are exponentially suppressed for UV localized ({\it i.e.} light) 
fermions. This is known as RS-GIM mechanism \cite{Agashe:2004ay,Agashe:2004cp}.
Details about the couplings and overlaps within the minimal RS formulation, with an IR-brane 
Higgs as well as gauge- and matter fields in the bulk, can be found in \cite{Casagrande:2008hr}.
The famous custodial extension including a protection for the $Zb_L\bar b_L$ vertex 
\cite{Agashe:2003zs,Agashe:2006at} is treated in \cite{Albrecht:2009xr,Casagrande:2010si}.

If one deals with SM-like quarks, it is convenient to expand the profiles 
in terms of $v^2/\Mkk^2$, where $v\approx 246$ GeV is the Higgs vacuum expectation value. 
This involves the zero-mode profile evaluated at the IR-brane  
\beq \label{eq:ZMP}
   F(c) = \mbox{sgn}[\cos(\pi c)]\, 
   \sqrt{\frac{1+2c}{1-\epsilon^{1+2c}}}
\eeq
as a function of the bulk mass parameters $c_{Q_i}=M_{Q_i}/k$ and $c_{q_i}=-M_{q_i}/k\,$\cite{Casagrande:2008hr}. 
To leading order (LO) in $v^2/\Mkk^2$ the spectrum of the light down-type quarks corresponds 
to the eigenvalues of the effective Yukawa matrix
\begin{align}\label{Yeff}
\begin{split}\displaystyle
   \bm{Y}_d^{\rm eff} =&\, \mbox{diag}\left[ F(c_{Q_i}) \right]
   \bm{Y}_d\,\mbox{diag}\left[ F(c_{d_i}) \right]\\
   =&\, \frac{\sqrt 2}v\, \bm{U}_d\,\mbox{diag}\left[m_d,m_s,m_b\right]\bm{W}_d^\dagger \,.
\end{split}
\end{align}
The mixing matrices $\bm{U}_d$ and $\bm{W}_d$ are most easily obtained by a singular-value 
decomposition of the left-hand side of the latter equality.
From now on, we will refer to the first non-trivial order in the expansion in $v^2/\Mkk^2$ (which 
we also apply for massive gauge bosons) as the zero-mode approximation (ZMA).

\section{Calculation of $\bm \Gamma_{12}^s$}

Within the SM, $\Gamma_{12}^s$ is known to next-to-leading order (NLO) in QCD  
\cite{Beneke:1996gn,Beneke:1998sy,Dighe:2001gc,Beneke:2003az,Ciuchini:2003ww,Badin:2007bv,Lenz:2006hd}.
In this section, we calculate the leading contribution to $\Gamma_{12}^s$ in the 
presence of NP. It is given by the hadronic matrix element of the transition amplitude, 
which converts $\bar B^0_s$ into $B^0_s$
\begin{align}
\begin{split}
\displaystyle
\Gamma_{12}^s=&\;\frac 1{2m_{B_s}}\, \langle B_s^0| {\mathcal T} | \bar B_s^0 \rangle\,,\\
{\mathcal T}=&\;\text{Disc} \int d^4x\,\frac i2\,T\,\left[
{\mathcal H}_{\rm eff}^{\Delta B=1}(x) {\mathcal H}_{\rm eff}^{\Delta B=1}(0)\right]\,.
\end{split}
\end{align}
Taking the discontinuity in the expression above projects out those intermediate states, 
that are on-shell. The leading correction to the SM result is given by the interference 
between SM and NP insertions.
The framework of heavy-quark expansion (HQE) allows for a systematic evaluation of
the matrix element in powers of $1/m_b$. At the zeroth order, the momentum of the $B$-meson
in its rest frame corresponds to the momentum of the bottom quark, while the strange-quark 
momentum is set to zero. At typical hadronic distances $x>1/m_b$, the transition of
$\bar B^0_s$ into $B^0_s$ is a local process. Thus, the matrix element can be expanded
in terms of local $\Delta B=2$ operators. QCD corrections are implemented by running the 
$\Delta B=1$ operators from the matching scale down to the mass of the bottom quark.
The leading SM contributions can be collected into matrix elements of the $\Delta B=2$ oper\-ators
\begin{align}\label{eq:Q1Q2}
\begin{split}
\Q_1=&\,(\bar s_i b_i)_{\small V-A}(\bar s_j b_j)_{\small V-A}\,,\\
\Q_2=&\,(\bar s_i b_i)_{\small S+P}(\bar s_j b_j)_{\small S+P}\,,
\end{split}
\end{align}
where $i$ and $j$ denote color indices and a summation over repeated indices is always 
understood throughout this article. The shorthand notation $V\pm A$ indicates the Dirac 
structure $\gamma^\mu(1 \pm \gamma^5)$ in between the spinors, whereas $S\pm P$ denotes 
$(1\pm \gamma^5)$.  
The possibility of having right-handed charged currents within the RS model asks for 
further $\Delta B=2$ operators, caused by interference of SM with NP insertions.
We introduce
\begin{align}\label{eq:Q3Q4Q5}
\begin{split}
\Q_3=&\,(\bar s_i b_j)_{\small S+P}(\bar s_j b_i)_{\small S+P}\,,\\
\Q_4=&\,(\bar s_i b_i)_{\small S-P}(\bar s_j b_j)_{\small S+P}\,,\\
\Q_5=&\,(\bar s_i b_j)_{\small S-P}(\bar s_j b_i)_{\small S+P}\,.
\end{split}
\end{align}

The appropriate $\Delta B=1$ Hamiltonian, allowing for new right-handed charged currents
as well as FCNCs, is given by 
\begin{widetext} 
\beq\label{eq:Heff}
\displaystyle
{\mathcal H}_{\rm eff}^{\Delta B=1}=\,\frac {G_F}{\sqrt 2}\,\lambda_{c}^{bs}
\,\Big{[}\sum_{i=1,2}\Big{(}C_iQ_i  \,+ \,C_i^{LL}Q_i\,
+\,C_i^{LR}Q_i^{LR}\,+\, C_i^{RL}Q_i^{RL}\Big{)}
+\sum_{i=3}^{10} C_i Q_i \Big{]}
+\sum_{i=3}^{10}\left(C_i^{\rm NP} Q_i + \widetilde C_i^{\rm NP}\widetilde Q_i\right).
\eeq
\end{widetext}
In the RS model the operators $Q_{1,2}$ arise from (KK) $W^\pm$-boson exchange, and the $LR/RL$ 
operators involve right-handed charged currents. They are defined as
\begin{align}\label{eq:Qcharge}
\begin{split}
Q_1&=(\bar s_i c_j)_{\small V-A}(\bar c_j b_i)_{\small V-A}\,,\\
Q_2&=(\bar s_i c_i)_{\small V-A}(\bar c_j b_j)_{\small V-A}\,,\\
Q_1^{LR}&=(\bar s_i c_j)_{\small V-A}(\bar c_j b_i)_{\small V+A}\,,\\
Q_2^{LR}&=(\bar s_i c_i)_{\small V-A}(\bar c_j b_j)_{\small V+A}\,,\\
\end{split}
\end{align}
and the $Q_i^{RL}$ are chirality flipped with respect to $Q_i^{LR}\,$. 
Operators of the type $RR$ are not included into our analy\-sis
as their coefficients scale like $v^4/\Mkk^4$ in the models at hand. 
Due to the hierarchies in the Cabibbo-Kobayashi-Maskawa (CKM) matrix and
the RS-GIM mechanism, it is sufficient to restrict ourselves on $c$ quarks 
as intermediate states, when we calculate the RS corrections
involving the charged current-sector. For the SM contribution however, we include
the combinations $uc$, $cu$, and $uu$ in addition to the operators given above.
Concerning the NP corrections $LL,LR,RL$, we pull out the CKM factor 
$\lambda_{c}^{bs}$ for conven\-ience. 
The measured values for $V_{cb}$ and $V_{cs}$, extracted from semileptonic $B$ and $D$ decays, 
should be identified with the exchange of all ($SU(2)_L$) $W$-type bosons.
As a consequence, the NP coefficients $C_{1,2}^{LL}$ arise only due to non-factoriz\-able corrections, 
which can not be absorbed into $\lambda_{c}^{bs}$. 
We further have to include QCD penguin opera\-tors
\begin{align}\label{eq:QCDpeng}
\begin{split}
Q_3=&\, (\bar s_i b_i)_{\small V-A} {\sum}_q\,(\bar q_j q_j)_{\small V-A}\,,\\
Q_4=&\, (\bar s_i b_j)_{\small V-A} {\sum}_q\,(\bar q_j q_i)_{\small V-A}\,,\\ 
Q_5=&\, (\bar s_i b_i)_{\small V-A} {\sum}_q\,(\bar q_j q_j)_{\small V+A}\,,\\
Q_6=&\, (\bar s_i b_j)_{\small V-A} {\sum}_q\,(\bar q_j q_i)_{\small V+A}\,,
\end{split}
\end{align}
as well as EW penguin operators
\begin{align}\label{eq:EWpeng}
\begin{split}
\displaystyle
Q_7=&\,\frac 32\,(\bar s_i b_i)_{\small V-A} {\sum}_q Q_q\,(\bar q_j q_j)_{\small V+A}\,,\\
Q_8=&\,\frac 32\,(\bar s_i b_j)_{\small V-A} {\sum}_q Q_q\,(\bar q_j q_i)_{\small V+A}\,,\\
Q_9=&\,\frac 32\,(\bar s_i b_i)_{\small V-A} {\sum}_q Q_q\,(\bar q_j q_j)_{\small V-A}\,,\\
Q_{10}=&\,\frac 32\,(\bar s_i b_j)_{\small V-A} {\sum}_q Q_q\,(\bar q_j q_i)_{\small V-A}\,,
\end{split}
\end{align}
where $q=u,c,d,s$, and $Q_q$ is the electric charge.
Here, no CKM factors are involved and one has to keep all
light quarks as intermediate states if one considers neutral-current insertions only.
The operators $\widetilde Q_{3..10}$ are chirality-flipped with respect to (\ref{eq:QCDpeng}) and 
(\ref{eq:EWpeng}). 
In principle, there is the possibility of a flavor change on both vertices for NP penguins
and the Wilson coefficients depend on the quark flavor $q$. 
However, these effects suffer from an additional RS-GIM suppression and 
can be neglected for all practical purposes.
For the same reason the chirality flipped penguins $\widetilde C_{3-10}^{\rm RS}$ can be neglected
compared to $C_{3-10}^{\rm RS}$ for $bs$ transitions \cite{Bauer:2009cf}. 
Within the minimal RS model it will turn out that, despite of the $\alpha/\alpha_s$-suppression, 
the EW penguin operators can dominate over the gluon penguins \cite{Agashe:2004ay,Agashe:2004cp,Bauer:2009cf}. 
This is explained by an extra factor $L$, which shows up in the leading 
correction to the left-handed $Z^0$-coupling. Note that this is not the case in the custodial 
RS variant \cite{Agashe:2003zs,Agashe:2006at}, which features a protection for the 
$Zb_L\bar b_L$ vertex. 
The RS Wilson coefficients of the penguin operators can be found in \cite{Bauer:2009cf}
and are collected in Appendix \ref{app:WilsonsP} for completeness.   
There further is the possibility of flavor-changing Higgs couplings which,
however, can be neglected against the contributions of flavor-changing heavy gauge bosons in RS models.

Concerning the double-penguin insertions, we include all light quarks with masses set to zero (besides $m_c$).
The double-penguin insertion also allows for leptons within the cut-diagram. However, as the related SM coefficient is suppressed 
by $\alpha/\alpha_s$, there is no chance to obtain big effects from $\bar sb\rightarrow\bar \tau \tau$
transitions, which are less constrained by experiment \cite{Bauer:2010dga}. 
Note that this is not a general statement about NP models. If there is a tree-level
transition $\bar sb\rightarrow\bar \tau \tau$ mediated by light NP particles in the range of 
$\sim 100\,$GeV, the double NP insertion becomes comparable to the SM diagrams \cite{Alok:2010ij}. 
Possible candidates are scalar leptoquarks \cite{Dighe:2007gt,Dighe:2010nj}.
Neglecting intermediate leptons, we find to LO in $1/m_b$
\begin{widetext} 
\begin{align}\label{eq:gamma12}
\displaystyle
\Gamma_{12}^s=-&\,\frac{m_b^2}{12\pi (2 M_{B_s})} \,G_F^2{(\lambda_{c}^{bs})}^2\,\sqrt{1-4z}\nr\\
\Bigg{\lbrace}&\Big{[}(1-z)(\Sigma_1+\Sigma_1^{LL})
  +\frac 12(1-4z)(\Sigma_2+\Sigma_2^{LL})\, +3z \,(\Sigma_3\,+K_3^{' LL})
  -\,\frac 32\,\sqrt z\, (\Sigma_1^{LR} +\Sigma_2^{LR}+K_3^{'LR}+K_4^{'LR})\nr\\
&\  +\frac 1{\sqrt{1-4z}}\Big{(}(3\bar K''_1 +K''_{s1} +\frac 32 \bar K''_2 + \frac 12 K''_{s2})
  \, +\, \frac{\lambda_u^{bs}}{\lambda_c^{bs}}\,(1-z)^2\big{(}(2+z)K_1+(1-z)K_2\big{)}
     +\frac 12 \frac{{(\lambda_u^{bs})}^2}{{(\lambda_c^{bs})}^2}\,(2K_1+K_2) \Big{)}\Big{]}\,
           \langle{\Q_1}\rangle\nr\\
 +&\, \Big{[}(1+2z)(\,\Sigma_1+\Sigma_1^{LL}-\,\Sigma_2-\Sigma_2^{LL})
  \,-\,3\,\sqrt z\,(2\Sigma_1^{LR} +\Sigma_2^{LR} -K_4^{'LR})\nr \\
&\  +\frac 1{\sqrt{1-4z}}\Big{(}(3\bar K''_1 +K''_{s1}-3\bar K''_2-K''_{s2})
    +\,2\, \frac{\lambda_u^{bs}}{\lambda_c^{bs}}\,(1-z)^2(1+2z)(K_1-K_2)
           +\,\frac{{(\lambda_u^{bs})}^2}{{(\lambda_c^{bs})}^2}(K_1-K_2)\Big{)}  \Big{]}\,
   \langle{\Q_2}\rangle \nr\\
 -&\;3\,\sqrt z\,(\Sigma_1^{LR} +2\Sigma_2^{LR}+K_3^{'LR})\,\langle \Q_3 \rangle \;
   +\, 3\sqrt z\,(\Sigma_1^{RL}-K_3^{'RL})\,\langle\Q_4\rangle
\; +\,3\sqrt z\,(\Sigma_2^{RL}-K_4^{'RL})\,\langle\Q_5\rangle \,\Bigg{\rbrace} \nr\\
-&\frac{m_b^2}{12\pi (2 M_{B_s})}\,\sqrt 2\,G_F\lambda_{c}^{bs}\,\sqrt{1-4z}\nr\\
\Bigg{\lbrace}&\Big{[}(1-z)\,\Sigma_1^{\rm NP}
+\frac 12(1-4z)\,\Sigma_2^{\rm NP} +\,3z \,\Sigma_3^{\rm NP}
+ \frac 1{\sqrt{1-4z}}(3\bar K_1^{''\rm NP} +K_{s1}^{''\rm NP}
+ \frac 32 \bar K_2^{''\rm NP} + \frac 12 K_{s2}^{''\rm NP})\Big{]}\,\langle{\Q_1}\rangle\\
&\,+\Big{[}(1+2z)(\,\Sigma_1^{\rm NP}-\,\Sigma_2^{\rm NP})
 + \frac 1{\sqrt{1-4z}}(3\bar K_1^{''\rm NP}+K_{s1}^{''\rm NP}
 -3\bar K_2^{''\rm NP}-K_{s2}^{''\rm NP})\Big{]}\,\langle{\Q_2}\rangle
\,+\;\ord\left(\frac 1{m_b}\right)\Bigg{\rbrace}\,,\nr
\end{align}
\end{widetext}
where $z=m_c^2/m_b^2\,$ and 
$\langle\Q\rangle\,\equiv\,\langle B_s^0|\, \Q\,|\bar B_s^0\rangle$.
In order to get a compact result, we have defined the linear combinations 
($A,B \in \lbrace L,R \rbrace$)
\begin{align}\label{eq:coefsum}
\begin{split}
\Sigma_i=&\,K_i+K'_i+K''_i\, , \\
\Sigma_i^{AB}=&\,K_i^{AB}+K_i^{' AB},\ \ i=1,2,\\
\Sigma_3=&\,K'_3+K''_3\,,\\
\Sigma_i^{\rm NP}=&\,K_i^{'\rm NP}+K_i^{''\rm NP}\,\ \ i=1,2,3\,,
\end{split}
\end{align}
where the coefficients on the right-hand side of (\ref{eq:coefsum}) are themselves 
linear combinations of Wilson coefficients. 
In agreement with \cite{Beneke:1996gn} we have ($C_{i+j}\equiv C_i+C_j$)
\begin{align}
K_1=&\,N_cC_1^2+2\,C_1C_2\,,\quad K_2=\,C_2^2\,,\nr\\
K'_1=&\,2\,(N_cC_1C_{3+9}+C_1C_{4+10}+C_2C_{3+9})\,,\nr\\
K'_2=&\,2\,C_2C_{4+10}\,,\nr\\
K'_3=&\,2\,(N_cC_1C_{5+7}+C_1C_{6+8}+C_2C_{5+7}+C_2C_{6+8})\,,\nr\\
K''_1=&\,N_cC_{3+9}^2+2\,C_{3+9}C_{4+10}\nr\\
      &\ \ +N_cC_{5+7}^2+2\,C_{5+7}C_{6+8}\,,\\
K''_2=&\,C_{4+10}^2+C_{6+8}^2\,,\nr\\ 
K''_3=&\,2(N_c C_{3+9}C_{5+7}+C_{3+9}C_{6+8}\nr\\
&\quad+C_{4+10}C_{5+7}+C_{4+10}C_{6+8})\,.\nr
\end{align} 
The combinations $K_i$ stem from the insertion of charged-current operators
and give the dominant contribution in the SM.
The coefficients $K'_i$ and $K''_i$ correspond to the interference of charged-current
with penguin operators and penguin-penguin insertions, respectively.
As we consider light quarks ($q=u,d,s\,$) in
the limit $m_q=0$, there is a cancellation in the EW penguin sector
due to the electric charges. The coefficients $\bar K_i^{''}$ therefore resemble 
the $K_i^{''}$, with $C_{7..10}$ set to zero.
For strange quarks as intermediate states, there is a second possibility for
the penguin insertion. In the limit $m_s=0$, there are additional contributions from
\begin{align}
K''_{s1}=&\,(2+N_c)(C_4- C_{10}/2)^2 \\ 
	&\ \ + 2\,(N_c+1)(C_3-C_9/2)(C_4-C_{10}/2)\nr \\ 
	&\ \ + 2\,(C_3-C_9/2)^2\,,\nr\\
K''_{s2}=&\, 2\,(C_3-C_9/2)(C_4-C_{10}/2) + (C_3-C_9/2)^2 \,.\nr
\end{align}
Note that these terms have not been taken into account in \cite{Beneke:1996gn}.
However, as all double-penguin insertions are numerically suppressed, this omission 
has no significant effect.
Next we come to the interference of SM diagrams with NP penguins, which is 
collected in
\begin{align}
K_1^{'\rm NP}=&\,2\,(N_cC_1C_{3+9}^{\rm NP}+C_1C_{4+10}^{\rm NP}+C_2C_{3+9}^{\rm NP})\,,\nr\\
K_2^{'\rm NP}=&\,2\,C_2C_{4+10}^{\rm NP}\,,\nr\\
K_3^{'\rm NP}=&\,2\,(N_cC_1C_{5+7}^{\rm NP}+C_1C_{6+8}^{\rm NP}+C_2C_{5+7}^{\rm NP}+C_2C_{6+8}^{\rm NP})\,,\nr\\
K_{s1}^{''\rm NP}=&\,2\,\big{(}(N_c+2) C_4(C_4^{\rm NP}-C_{10}^{\rm NP}/2)\nr\\
                  &\ \ + (N_c+1) C_4(C_3^{\rm NP}-C_9^{\rm NP}/2)\\
                 &\ \ + (N_c+1) C_3(C_4^{\rm NP}-C_{10}^{\rm NP}/2)\nr\\
		&\ \ + 2 C_3(C_3^{\rm NP}-C_9^{\rm NP}/2) \big{)}\,,\nr\\
K_{s2}^{''\rm NP}=&\,2\,\big{(}C_3(C_3^{\rm NP}-C_9^{\rm NP}/2)+C_3(C_4^{\rm NP}-C_{10}^{\rm NP}/2)\nr\\
			&\ \ +C_4(C_3^{\rm NP}-C_9^{\rm NP}/2)\big{)}\,\nr
\end{align} 
and 
\begin{align}
K_1^{''\rm NP}=&\,2\,(N_cC_3C_{3+9}^{\rm NP}+C_3C_{4+10}^{\rm NP}+C_4C_{3+9}^{\rm NP}\nr\\
&\,+N_cC_5C_{5+7}^{\rm NP}+C_5C_{6+8}^{\rm NP}+C_6C_{5+7}^{\rm NP})\,,\\
K_2^{''\rm NP}=&\,2\,(C_4C_{4+10}^{\rm NP}+C_6C_{6+8}^{\rm NP})\,,\nr\\
K_3^{''\rm NP}=&\,2\,(N_c C_3C_{5+7}^{\rm NP}
               +C_3C_{6+8}^{\rm NP}+C_4C_{5+7}^{\rm NP}+C_4C_{6+8}^{\rm NP}\nr\\
               &\,+N_c C_5C_{3+9}^{\rm NP}+C_5C_{4+10}^{\rm NP}+C_6C_{3+9}^{\rm NP}+C_6C_{4+10}^{\rm NP})\,.\nr
\end{align}
Here, we have neglected the tiny contributions from the interference of SM EW penguins with NP graphs. 
There further is interference between NP charged currents and SM penguins
\begin{align}
K_1^{'LL}=&\,2\,(N_c C_3C_1^{LL}+C_3C_2^{LL}+C_4C_1^{LL})\,,\nr\\
K_2^{'LL}=&\,2\,C_4C_2^{LL} \,,\nr\\
K_3^{'LL}=&\,2\,(N_c C_5C_1^{LL}+C_5C_2^{LL}+C_6C_1^{LL}+C_6C_2^{LL})\,,\nr\\
K_1^{'LR}=&\,2\,(N_c C_3C_1^{LR}+C_3C_2^{LR}+C_4C_1^{LR})\,,\nr\\
K_2^{'LR}=&\,2\,C_4C_2^{LR}\,\nr\\
K_3^{'LR}=&\,2\,(N_c C_5C_1^{LR}+C_5C_2^{LR}+C_6C_1^{LR})\,,\nr\\
K_4^{'LR}=&\,2\,C_6C_2^{LR}\,.
\end{align}
The corrections to the purely charged-current interactions are collected into
\begin{align}
\begin{split}
K_1^{LL}=&\,2\,\big{(}N_cC_1C_1^{LL}+C_1C_2^{LL}+C_2C_1^{LL}\big{)}\,,\\\
K_2^{LL}=&\,2\,C_2C_2^{LL}\,,\\
K_1^{LR}=&\,2\,(N_cC_1C_1^{LR}+C_1C_2^{LR}+C_2C_1^{LR})\,,\\
K_2^{LR}=&\,2\,C_2C_2^{LR}\,.
\end{split}
\end{align}
The coefficients $K_i^{(')RL}$ resemble $K_i^{(')LR}$, with $C_i^{LR}$
replaced by $C_i^{RL}$.
All NP coefficients should by calculated at the NP mass scale and then be evolved
down to $m_b$. Explicit expressions for the minimal and the custodial RS model
can be found in the appendices. 

For the sake of completeness, we finally quote the known results for the mixing amplitude. 
One defines
\beq
 {\mathcal H}_{\rm eff}^{\Delta B=2}=\sum_{i=1}^5 C_i \Q_i+ \sum_{i=1}^3 \widetilde C_i \widetilde\Q_i\,,
\eeq
where there are no tree-level contributions to $C_{2,3}$ and $\widetilde C_{2,3}$ in the RS model 
\cite{Blanke:2008zb,Bauer:2009cf}. The RS correction to 
\beq\label{eq:M12me}
2\, m_{B_s} M_{12}^s=\, \langle B_s^0|\, {\mathcal H}_{\rm eff}^{\Delta B=2}\, | \bar B_s^0 \rangle\,
\eeq
can be found in \cite{Blanke:2008zb,Bauer:2009cf}, and is given by
\begin{align}\label{eq:M12RS}
M_{12}^{s\,{\rm RS}}=\, \frac 43\, m_{B_s}  f_{B_s}^2&\Big{[} 
\left(C_1^{\rm RS}(\bar m_b)+\widetilde C_1^{\rm RS}(\bar m_b)\right) B_1 \nr\\
&+\frac 34\, R(\bar m_b)\,C_4^{\rm RS}(\bar m_b)B_4\\
&+\frac 14\,R(\bar m_b)\,C_5^{\rm RS}(\bar m_b)B_5\Big{]}\,.\nr
\end{align}
The bag parameters $B_{1,4,5}$ are listed in (\ref{eq:bagfac}), and the $\Delta B=2$ coefficients 
can be found in Appendix \ref{sec:mixing}.
Compared to $C_1^{\rm RS}(\bar m_b)$, the coefficient $C_4^{\rm RS}(\bar m_b)$ is suppressed by 
about two orders of magnitude due to a stronger RS-GIM mechanism. The coefficients 
$\widetilde C_1^{\rm RS}(\bar m_b)$ and $C_5^{\rm RS}(\bar m_b)$ are even further suppressed.  
The SM mixing amplitude can be taken from \cite{Buras:1990fn,Buras:1998raa,Lenz:2006hd}
\beq\label{eq:M12SM}
\displaystyle
M_{12}^{s\,{\rm SM}}=\frac{G_F^2}{12\,\pi^2}\,
{(\lambda^{bs}_t)}^2 m_W^2 m_{B_s} \eta_B f_{B_s}^2 B_1 S_0(x_t)\,,
\eeq
where $\eta_B=0.837$ involves NLO QCD corrections in naive dimensional reduction (NDR).
$S_0(x_t)$ is the Inami-Lim function and $x_t=\bar m_t(\bar m_t)^2/m_W^2$ with $\bar m_t(\bar m_t)=(163.8\pm2.0)\,$GeV. 
The meson mass and decay constant are given by $m_{B_s}=5.366(1)\,$GeV \cite{Nakamura:2010zzi} and 
$f_{B_s}=(238.8\pm9.5)\,$MeV \cite{Laiho:2009eu}, respectively.
If not stated otherwise, all other experimental input is taken from \cite{Nakamura:2010zzi}.

\section{numerical analysis}\label{sec:num}

In order to obtain the RS predictions, we need an appropriate set of input parameters,
consisting of the Yukawa matrices, the bulk-mass parameters $c_{Q_i}$ and $c_{q_i}$ of the
$SU(2)_L$-doublet and singlet fermions, as well as the KK scale $\Mkk$. Within an anarchic
approach to flavor, all Yukawa entries are chosen to be of $\ord(1)$.
The generation of input points is most easily achieved by making use of the warped-space 
Froggatt-Nielsen mechanism \cite{Agashe:2004cp,Casagrande:2008hr}, which provides simple analytic 
expressions for the fermion masses and Wolfenstein parameters in terms of the zero-mode profiles (\ref{eq:ZMP})
and entries of the Yukawa matrices, but independent of $\Mkk$ to first approximation.
In our analysis, we use 10000 randomly generated parameter sets with $|(Y_{u,d})_{ij}|\in[0.1,3]$, which guarantees 
perturbativity of the Yukawa couplings in higher order corrections \cite{Csaki:2008zd}.  
The points are chosen such that they fit the correct 
zero-mode masses, CKM mixing angles and phase in standard convention 
\cite{Nakamura:2010zzi} within the $1\, \sigma$ range.
 
The contributions of some individual ingredients of $\Gamma_{12}^s$ 
(\ref{eq:gamma12}) are summarized in Table \ref{tab:coef}. The SM coefficients 
are taken from \cite{Buchalla:1995vs}. 
For the sake of comparison, we rescale the RS penguin coefficients, for instance 
$\tilde K_2^{'\rm RS}\equiv \sqrt 2\,{(G_F \lambda_c^{bs})}^{-1} K_2^{'\rm RS}$ 
(SM: $\tilde K'_2=K'_2$), as they are not supplemented with a CKM factor in (\ref{eq:Heff}).
We compare the mean absolute values of our RS predictions to the corresponding sizes of 
the SM coefficients, where the numbers have to be multiplied by the order of magnitude 
given in the last column. 
The maximum values exceed the given numbers by at least one order of magnitude, 
as suggested by the large standard deviations.
The NP mass scale is set to $\Mkk=2\,$TeV and we discard all points, which are in conflict
with the $Z^0\rightarrow b\bar b$ ``pseudo observables''. These are the ratio of the width 
of the $Z^0$-boson decay into bottom quarks and the total hadronic width, $R_b^0$, 
the bottom quark left-right asymmetry parameter $A_b$, and the forward-backward 
asymmetry for bottom quarks $A_{\rm FB}^{0,b}$,
which set an upper limit on $c_{b_L}\equiv c_{Q_3}$ \cite{Casagrande:2008hr}. 
For $\Mkk=2\,$TeV, most points with $c_{b_L}>-0.5$ are excluded. On the other hand, 
for $\ord(1)$ Yukawa couplings, the top-quark mass only allows for a minimal UV localization
of the $(t_L,b_L)^T$ doublet. Thus, the valid bulk-mass parameters $c_{Q_3}$ are clustered around $-1/2\,$. 
We reject all points which lie outside the $95\%$ confidence region 
in the $g_L^b-g_R^b$ plane (see analysis in \cite {Casagrande:2008hr}).
Within the custodial RS variant with protection of the $Z^0b_L\bar b_L$-vertex, the related 
upper bound on $c_{b_L}$ vanishes. 
On the other hand, there is no stringent upper bound on the bulk-mass parameter $c_{t_R}\equiv c_{u_3}$,
which we allow to vary within $[-0.5,1]$.

\begin{table}
\begin{center}
\begin{tabular}{|c|c|c|c|c|c|c|}
  \hline
  Model\slash Coef.  & $\, |\tilde K'_2|\,$ & $\, |\tilde K''_2|\,$  
                     & $|K_2^{(LL)}|$ & $|K_2^{LR}|$ & $|K_2^{RL}|$ & $\times$ \\
  \hline
  SM  & $0.543$ & $0.016$ & $12.656$ & - & - & $10^{-1}$ \\
  \hline
  mean(min RS)  & $0.16$ & $0.03$ & $0.01$ & $4.40$ & $0.04$ & $10^{-3}$\\
  stand. dev.  & $0.17$ & $0.03$ & $0.05$ & $7.41$ & $0.06$ & $10^{-3}$\\
  \hline
  mean(cust)  & $0.94$ & $0.06$ & $0.23$ & $2.22$ & $0.03$ & $10^{-3}$ \\
  stand. dev.  & $1.39$ & $0.09$ & $1.38$ & $4.98$ & $0.05$ & $10^{-3}$ \\
  \hline
\end{tabular}
\end{center}
\caption{\label{tab:coef}  Selected SM penguin and charged
  current coefficients contributing to $\Gamma_{12}^s$ compared to 
  the mean absolute values of the corresponding RS coefficients for 
  $\Mkk=2\,$TeV and $\mu=\bar m_b\,$. See text for details. }
\end{table}

Neglecting experimental constraints, there is no difference between the minimal and the custodial 
RS variant at LO in $v^2/\Mkk^2$ in the charged current sector (see Appendix \ref{app:charged}). 
For the natural assumption of $c_{Q_2}<-1/2\,$ the biggest correction comes from the operator $Q_2^{LR}$. 
This is easy to understand if we apply the Froggatt-Nielsen analysis of \cite{Casagrande:2008hr} 
to (\ref{eq:CLL}) and (\ref{eq:CLRCRL}). 
Setting all Yukawa factors to one, we can derive simple expressions for the Wilsons coefficients 
by performing an expansion in the Wolfenstein parameter
$\lambda\approx 0.225\,$, which is related to ratios of IR zero-mode profiles \cite{Casagrande:2008hr}
\beq
\displaystyle
\frac{|F(c_{Q_1})|}{|F(c_{Q_2})|}\sim\lambda\,,\quad
\frac{|F(c_{Q_2})|}{|F(c_{Q_3})|}\sim\lambda^2\,,\quad |F(c_{Q_3})|\sim\ord(1)\,.
\eeq
Thus, we find as a crude approximation
\begin{align}
\displaystyle
C_2^{LL}\propto\;& \frac{m_W^2}{2\Mkk^2}\, L\;F(c_{Q_2})^2 F(c_{Q_3})^2\,,\\
C_2^{LR}\propto\;& \frac{v^2}{2\Mkk^2}\,\frac{F(c_{Q_3})}{F(c_{Q_2})}\, F(c_{u_2}) F(c_{d_3})
\propto \frac{m_c m_b}{\Mkk^2}\,\frac 1{F(c_{Q_2})^2} \,,\nr\\
C_2^{RL}\propto\;&\frac{v^2}{\Mkk^2}\,F(c_{u_2}) F(c_{d_2})\propto \frac{2\,m_c m_s}{\Mkk^2}
\frac 1{F(c_{Q_2})^2} \nr\,.
\end{align}
Note that the importance of $C_2^{LR}$ grows with increasing UV-localization of the $(c_L,s_L)^T$ doublet.
The coefficients $C_1^{AB}$ with $A,B \in \lbrace L,R \rbrace$ are zero at the matching scale, but generated
through operator mixing when running down to $\mu=\bar m_b$.
As it turns out, the values of $|K_1^{AB}|$ are about a third of the 
respective values of  $|K_2^{AB}|$ at $\mu=\bar m_b$.
In the RS model the contributions from the coefficients $C_i^{LL}$ and $C_i^{RL}$ can be neglected, just as those
of the chirality flipped penguins.
The coefficients $K_i^{'{\rm RS}}$ and $K_i^{''{\rm RS}}$ grow with an increasing value of 
$c_{b_L}$ and $ c_{s_L} \equiv c_{Q_2}$.
The reason is that the RS corrections due to penguin operators are dominated by overlap 
integrals of left-handed fermions with intermediate
KK-gauge bosons and mixing effects of the latter with $Z^0$. The relevant expressions
are given in (\ref{eq:Cpenguin}). As KK modes are 
peaked towards the IR brane, overlap integrals with UV localized fermions are exponentially 
suppressed and RS-GIM is at work. The leading correction due to $Z^0$ exchange
is enhanced by a factor $L$ within the minimal RS variant.
Nevertheless, due to the stringent bounds from $Z^0b\bar b$, the total penguin contributions remain 
smaller than in the custodial model.
In both models, it is sufficient to consider just the contributions stemming from
the coefficients $K_i^{'{\rm NP}}$ in the neutral-current sector.
The impact of double penguins is typically about $1\%$ of the leading correction
due to charged currents. 

In order to get the overall picture, we have to evaluate the whole expressions (\ref{eq:gamma12})
and (\ref{eq:M12RS}).
In terms of
\begin{equation}
R(\mu)\equiv {\left(\frac{M_{B_s}}{\bar{m}_b(\mu)+\bar{m}_s(\mu)}\right)}^2\,,
\end{equation}
the matrix elements are given by
\begin{align}\label{eq:bag}
\begin{split}
\langle\Q_1\rangle&\,=\,\frac 83\, M_{B_s}^2 f_{B_s}^2 B_1(\mu)\,,\\
\langle\Q_2\rangle&\,=-\frac 53\, M_{B_s}^2 f_{B_s}^2 \,R(\mu)\, B_2(\mu)\,,\\
\langle\Q_3\rangle&\,=\,\frac 13\, M_{B_s}^2 f_{B_s}^2 R(\mu) B_3(\mu)\,,\\
\langle\Q_4\rangle&\,=\,2\, M_{B_s}^2 f_{B_s}^2 R(\mu) B_4(\mu)\,,\\
\langle\Q_5\rangle&\,=\,\frac 23\, M_{B_s}^2 f_{B_s}^2 R(\mu) B_5(\mu)\,.
\end{split}
\end{align}
The bag parameters $B_i$ can be extracted from the lattice. We take the values of
\cite{Becirevic:2001xt} in the NDR-$\overline{\rm MS}$ scheme of \cite{Beneke:1998sy}.
They read
\begin{align}\label{eq:bagfac}
& B_1=0.87(2)\left(^{+5}_{-4}\right),\ \ B_2=0.84(2)(4),\ \ B_3=0.91(3)(8),\nr \\
& B_4=1.16(2)\left(^{+5}_{-7}\right),\ \ B_5=\,1.75(3)\left(^{+21}_{-6}\right),
\end{align}
where the first (second) number in brackets corresponds to the statistical (systematic) error. 
In order to resum large logarithms we employ $\bar z=\bar m_c^2(\bar m_b)/\bar m_b^2(\bar m_b)=0.048(4)$ 
\cite{Lenz:2006hd} in our numerical analysis. We further use $\bar m_b(\bar m_b) = (4.22 \pm 0.08)$\,GeV 
and $\bar m_s(\bar m_b) = (0.085 \pm 0.017)$\,GeV. 

\begin{figure}[t!]
\begin{center}
\includegraphics[width=6.5cm]{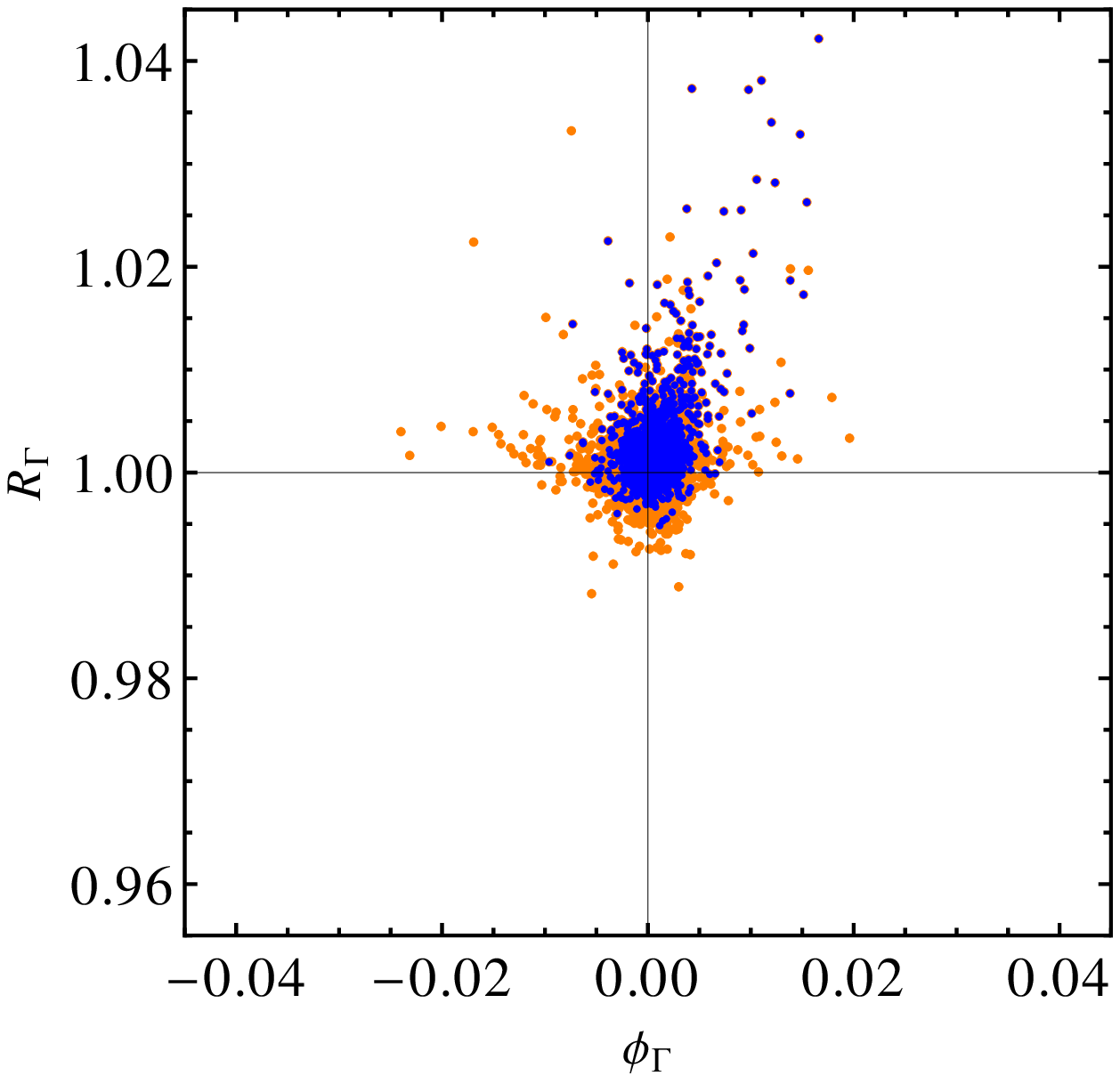}
\end{center}
\vspace{-2.5mm}
\begin{center}
\includegraphics[width=6.5cm]{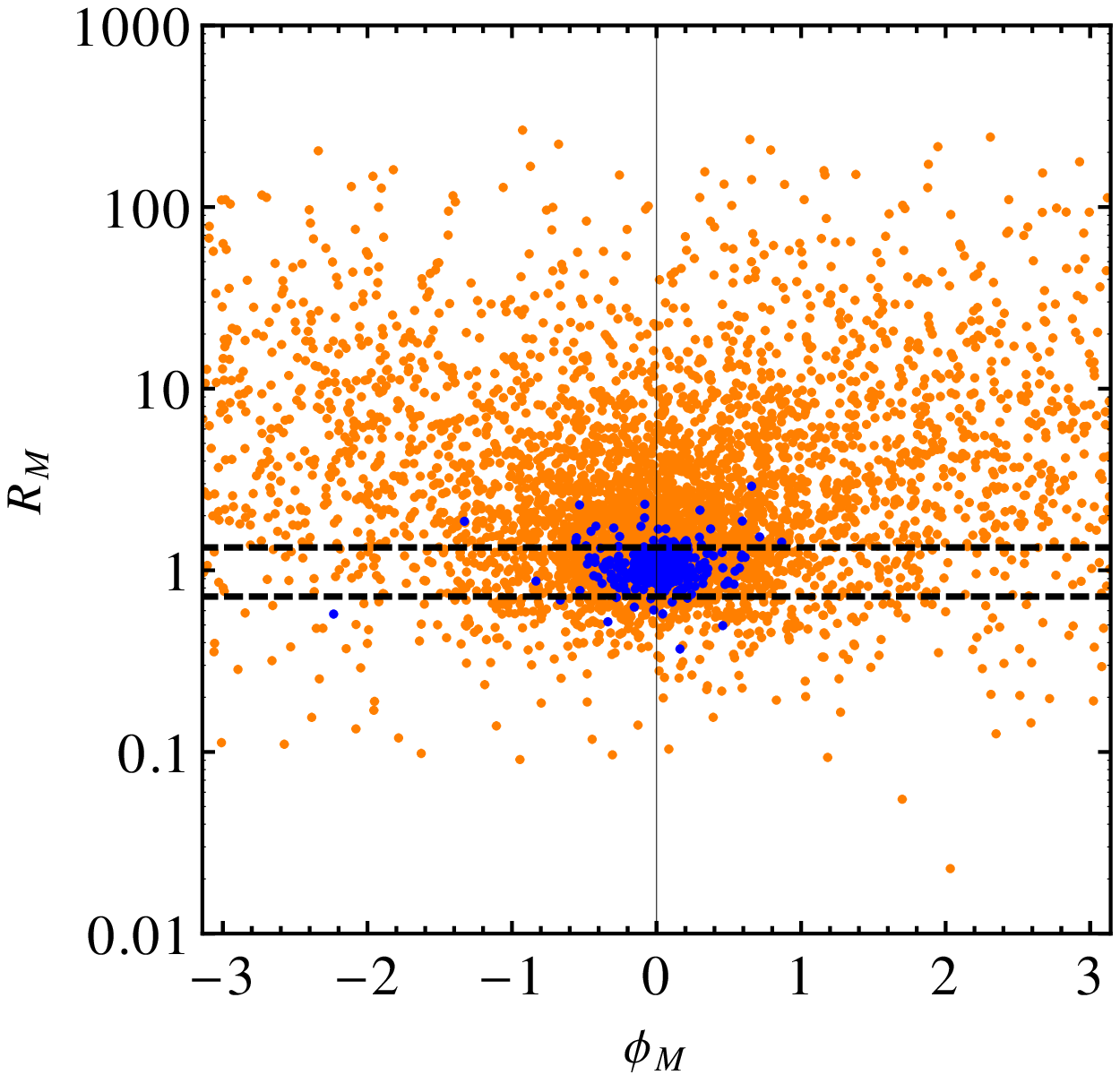}
\end{center}
\vspace{-7.5mm}
\begin{center}
  \parbox{8cm}{\caption{\label{fig:RS} RS corrections to the magnitude and CP-violating phase
  of the $\bar B_s^0$-$B_s^0$ decay amplitude, $R_\Gamma$ and $\phi_\Gamma$, as well as for the mixing amplitude, 
  $R_M$ an $\phi_M$. Blue (dark gray) points correspond to the minimal, orange (light gray) to the 
  custodial RS model. The dashed lines mark the $95\%$ confidence region
  with respect to the measurement of $\Delta m_{B_s}$. See text for details.}}
\end{center}
\end{figure}

In the first panel of Figure \ref{fig:RS} we show the RS corrections to the magnitude 
and CP-violating phase of the $\bar B_s^0$-$B_s^0$ decay width, $R_\Gamma$ and $\phi_\Gamma$, for a set 
of 10000 parameter points at $\Mkk=2\,$TeV. The blue (dark gray) points correspond 
to the minimal RS model, where we plot only those that are in agreement with the 
$Z^0\rightarrow b\bar b$ ``pseudo observables''. 
The orange (light gray) points correspond to the custodial extension, where the latter bound vanishes. 
As we are just interested in the approximate size of RS corrections, we work with the LO SM expressions. 
For precise predictions for a certain parameter point, one should include the full NLO corrections 
to $\Gamma_{12}^s$ and $M_{12}^s$.
As expected, the RS corrections to $|\Gamma_{12}^s|$ are rather small, typically not exceeding $\pm 4\%$.
The corrections to the magnitude and phase of the dispersive part of the mixing amplitude, 
$R_M$ and $\phi_M$, are plotted in the second panel of Figure \ref{fig:RS}. 
At this point, one should keep in mind the experimental result from the  
measurement of the $\bar B_s^0$-$B_s^0$ oscillation frequency \cite{Abulencia:2006ze}
\beq\label{eq:msexp}
\Delta m_{B_s}^{\rm exp}= (17.77\pm 0.10\, ({\rm stat})\pm0.07\,({\rm syst}))\,{\rm ps}^{-1}\, ,
\eeq
which is in good agreement with the SM prediction  
$(17.3\pm2.6)\,{\rm ps}^{-1}$ \cite{Lenz:2011ti}.
As a consequence, all points with $R_M \not\in [0.718,1.336]$ are excluded at $95\%$ confidence level, as indicated by the 
dashed lines.
For a sufficient amount of scatter points, the phase correction $\phi_M$  
can take any value of $[-\pi,\pi]$ within the custodial RS model.
Compared to $\phi_M$, the new phase $\phi_\Gamma$ can be neglected (what we will do from now on). 

\begin{figure}[t!]
\begin{center}
\includegraphics[width=6.7cm]{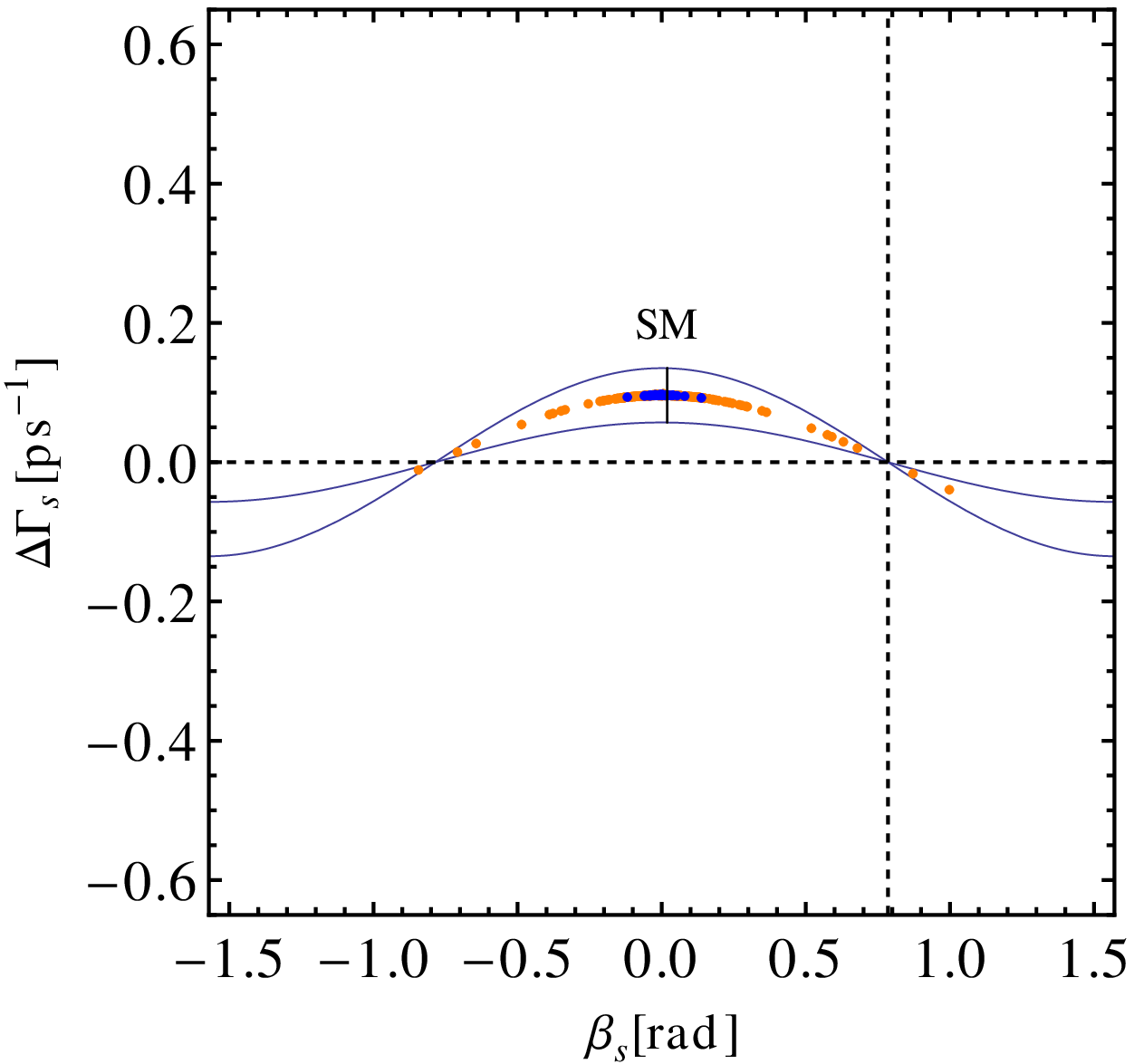}
\end{center}
\vspace{-2.5mm}
\begin{center}
\includegraphics[width=7.3cm]{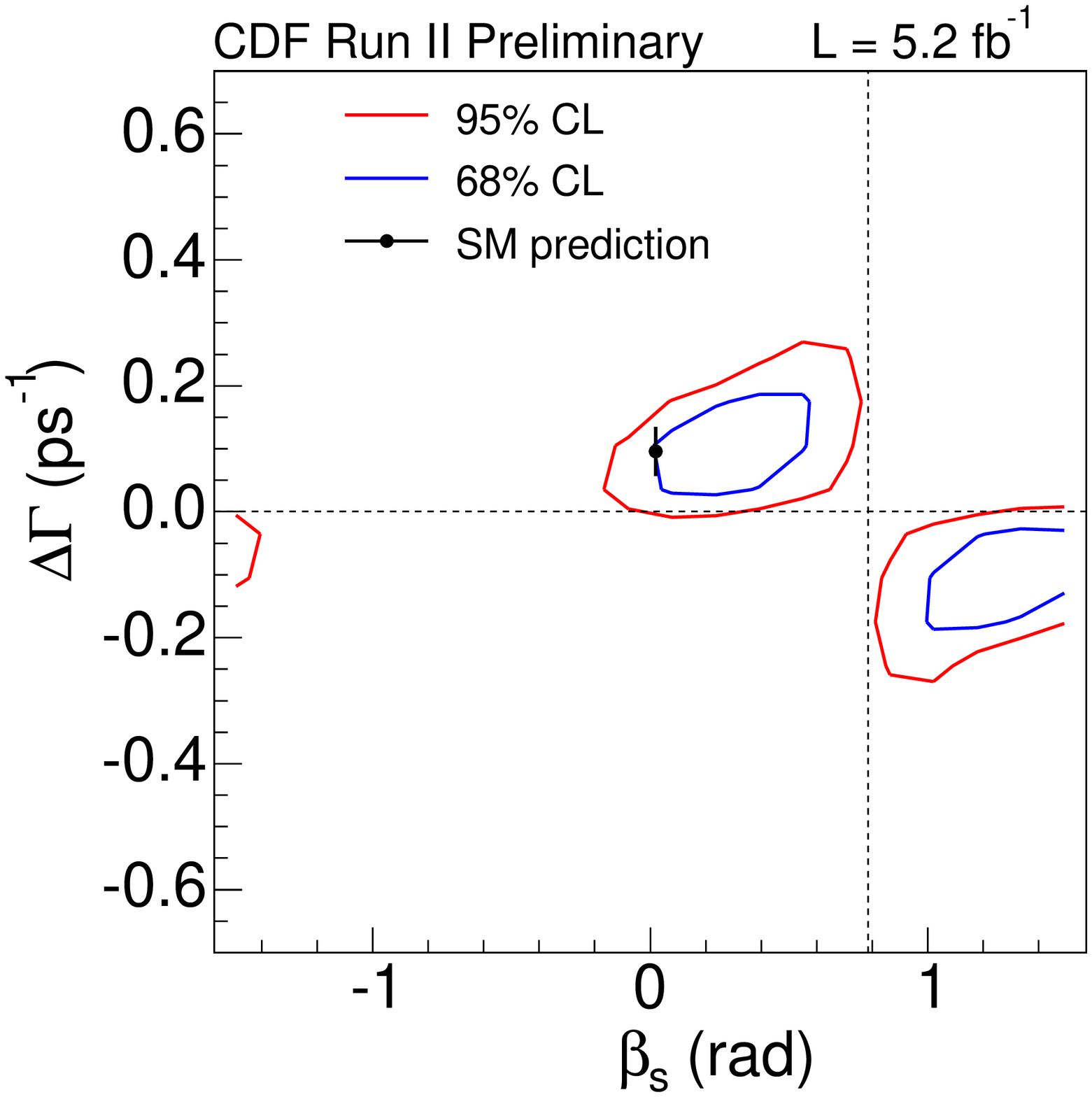}
\end{center}
\vspace{-2.5mm}
\begin{center}
  \parbox{8cm}{\caption{\label{fig:gammabeta} Upper panel: Corrections 
   within the $\Delta\Gamma_s^{\rm SM}/\beta_s\,$-plane for
   the minimal (blue/dark gray) and custodial (orange/light gray) RS model.
   Bounds from $Z^0b\bar b$, $\Delta m_{B_s}$, and $\eps_K$ are satisfied. See text for details.
   Lower panel: Experimental constraints from flavor-tagged
   $B\rightarrow_s^0 J/\psi\phi\,$ decays. Figure taken from \cite{pubn2}. }}
\end{center}
\end{figure}
 
\begin{figure}[t!]
\begin{center}
\includegraphics[width=6.7cm]{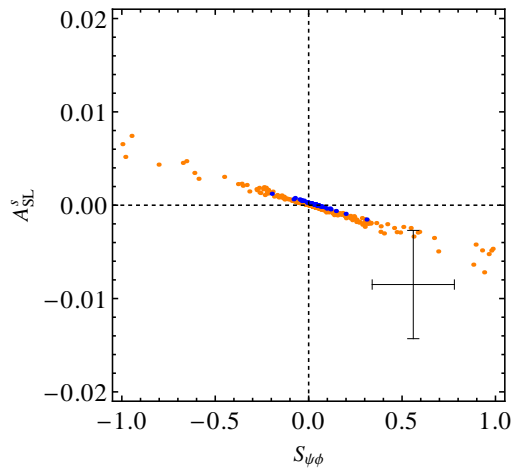}
\end{center}
\vspace{-2.5mm}
\begin{center}
  \parbox{8cm}{\caption{\label{fig:ASL} Corrections 
   within the $A_{\rm SL}^s/S_{\psi\phi}\,$-plane for
   the minimal (blue/dark gray) and custodial (orange/light gray) RS model.
   Bounds from $Z^0b\bar b$, $\Delta m_{B_s}$, and $\eps_K$ are satisfied. See text for details. }}
\end{center}
\end{figure}

We further take into account additional constraints from $\eps_K=\eps_K^{\rm SM}+\eps_K^{\rm RS}$ 
\cite{Blanke:2008zb,Bauer:2009cf,Csaki:2008zd,Davoudiasl:2008hx,Bauer:2008xb}. 
Explicitly, one needs to satisfy $|\eps_K|\in [1.2,3.2]\cdot 10^{-3}$, 
where 
\beq\displaystyle
\eps_K=\frac{\kappa_\eps\, e^{i\varphi_\eps}}{\sqrt 2\,{(\Delta m_K)}_{\rm exp}}\,
{\rm Im}({M_{12}^{K{\rm SM}}}+{M_{12}^{K{\rm RS}}})\,,
\eeq
with $\varphi_\eps=(43.51\pm0.05)^{\circ}$ \cite{Nakamura:2010zzi} and $\kappa_\eps=0.92\pm0.02$ 
\cite{Buras:2008nn}. The neutral kaon mixing amplitude is defined in analogy to (\ref{eq:M12me}). 
The input data needed for the calculation is given in Appendix {\it B} of \cite{Bauer:2009cf}. 
As it turns out, without some tuning, the prediction for $\eps_K$ is generically too large.
The dangerous contributions from the operators $Q_{4,5}^{sd}$ \cite{Csaki:2008zd,Davoudiasl:2008hx},
which can become comparable to those of $Q_1^{sd}$ due to $R_K=(M_K/(\bar m_d+\bar m_s))^2\approx 20\,$ for $\mu=2\,$GeV 
and a more pronounced RG running, can be suppressed 
by imposing a $U(3)$ flavor symmetry in the right-handed down-quark sector. This symmetry is 
broken by the Yukawa couplings to obtain the correct zero-mode masses \cite{Santiago:2008vq}. 
Nevertheless, if all bulk masses are equal, there are no tree-level FCNCs in the ZMA. 
This is evident from (\ref{eq:Cmix}), as ${(\bm{W}_d^\dagger)}_{mj}{(\bm{W}_d)}_{jn}=0\,$ for $m\neq n$ 
due to the unitarity of $\bm{W}_d\,$. Non-vanishing contributions from the exchange of KK gauge bosons arise 
from the mixing of the right-handed fermion zero modes with their KK excitations, 
thus involving an additional $v^2/\Mkk^2$-suppression factor.
For $\Mkk=2\,$TeV, we could therefore reduce $C_{4,5}^{sd}$ by a factor of about $100$. The same 
suppression factor then applies to the B-meson sector. For the coefficient $C_1^{\rm RS}$, there is no such protection. 
In our analysis however, we do not impose an additional flavor symmetry on the 
bulk masses, but rather use the bound from $\epsilon_K$ as a filter for our scattering points.

Neglecting the small SM phases, the width difference (\ref{eq:DGammaapprox}) can be written as
\beq\label{eq:Dgamma}
\Delta\Gamma_s= \Delta\Gamma_s^{\rm SM}\,R_\Gamma\,\cos 2\beta_s\,,
\eeq
where $2\beta_s\approx -\phi_M^{\rm RS}$ \cite{pubn}. The preliminary
CDF analysis \cite{pubn2} uses the older SM prediction 
$\Delta\Gamma_s^{\rm SM}=(0.096\pm0.039)\text{ps}^{-1}$ \cite{Lenz:2006hd},
which we will take as central value for our calculation.
Taking the more recent value will not change our conclusions.
The resulting RS predictions for $\Delta\Gamma_s$ are plotted against 
$\beta_s$ in the upper panel of Figure \ref{fig:gammabeta}. 
Comparing to the CDF results in the lower panel, we conclude that the 
RS model can enter the $68\%$ confidence region and come close to
the best fit value. It stays below the desired value for $\Delta\Gamma_s$, 
as there are no sizable positive corrections to $|\Gamma_{12}^s|$. 

It should be noted that the latest {\it LHCb} result of the phase 
$\phi_s^{J/\psi\phi}=-2\beta_s^{J/\psi\phi}=0.03\pm 0.16\pm 0.07$ agrees with
the SM prediction (\ref{eq:betasSM}) within errors. 
The above number combines measurements of $B_s^0$ decays 
into $J/\psi\phi\,$ and $J/\psi f_0\,$ \cite{LHCb:2011_1,LHCb:2011_2}.
In agreement with the Tevatron results, an enhancement of the width difference 
compared to the SM value has been found. 
The best-fit value is given by $\Delta\Gamma_s=(0.123\pm0.029\pm0.011){\rm ps}^{-1}$ \cite{LHCb:2011_1}.

The SM prediction ${(A_{\rm SL}^s)}_{\rm SM}=(1.9\pm0.3)\cdot 10^{-5}$ \cite{Lenz:2011ti}, which is 
often named $a_{\rm sl}^s$ or $a_{\rm fs}^s$ in the literature, agrees with the direct measurement 
${(A_{\rm SL}^s)}_{\rm exp}=-0.0017\pm0.0092$ \cite{Asner:2010qj} within the (large) error. 
However, recent measurements of the like-sign dimuon charge asymmetry $A_{\rm SL}^b$ \cite{Abazov:2010hv}, which 
connects $A_{\rm SL}^s$ to its counterpart $A_{\rm SL}^d$ of the $B_d^0$-meson sector 
\cite{Grossman:2006ce}, imply a deviation of almost $2\sigma$.
If one neglects the tiny SM phases and the NP phase corrections related to decay, $A_{\rm SL}^s$ is 
proportional to the quantity $S_{\psi\phi}$ \cite{Ligeti:2006pm}, which is given by the amplitude 
of the time-dependent asymmetry in $B_s^0\rightarrow J/\psi\phi$ decays, 
$A_{\rm CP}^s(t)=S_{\psi\phi}\sin(\Delta m_{B_s} t)$. Setting just the NP phase in decay to zero,
one obtains the well known expression $S_{\psi\phi}=\sin(2\beta_s^{J/\psi \phi}-\phi_M)$ \cite{Blanke:2006ig}, and thus
\beq
A_{\rm SL}^s\approx-\,\frac{|\Gamma_{12}^{s\,{\rm SM}}|}{|M_{12}^{s\,{\rm SM}}|}\,
\frac{R_\Gamma}{R_M}\,S_{\psi\phi}\,.
\eeq
The RS result is shown in Figure \ref{fig:ASL}, where  we have sketched the experimental favored values 
$S_{\psi\phi}=0.56\pm0.22$ \cite{Bona:2008jn} and $A_{\rm SL}^s=-0.0085\pm0.0058$ \cite{Asner:2010qj}. 
The latter number combines the direct measurement with the results derived from the measurement 
of $A_{\rm SL}^b$ in semileptonic $B$-decays together with the average 
$A_{\rm SL}^d=-0.0047\pm0.0046$ from $B$-factories.
It is evident from the plot that the best fit value of $S_{\psi\phi}$ can be reproduced 
(with some tuning in the minimal RS variant), which has already been noted in \cite{Blanke:2008zb,Bauer:2009cf}.
Furthermore, the custodial RS model can enter the $1\sigma$ range of the measured value of $A_{\rm SL}^s$.
The same conclusion has been drawn in \cite{Datta:2010yq} recently, using a different approach.
Here, the authors did not produce any concrete sets of input parameters, but scanned FCNC vertices
across the allowed range subject to bounds from $\Delta \Gamma_s$ and $\Delta m_{B_s}$.

Note that due to $S_{\psi\phi}\approx\sin 2\beta_s$, the corrections in the $\Delta\Gamma_s^{\rm SM}/\beta_s\,$-plane
and the $A_{\rm SL}^s/S_{\psi\phi}\,$-plane are correlated. An improvement in the former leads to an improvement
in the latter.

\section{Conclusions}\label{sec:conc}

In this article, we investigated the impact of RS models on the width difference $\Delta\Gamma_s$ 
of the $\bar B_s^0$-$ B_s^0$ system and the related CP violating observables $A_{SL}^s$ and $S_{\psi\phi}$.
Therefore we calculated the leading corrections to $\Gamma_{12}^s$ in terms 
of NP Wilson coefficients and took the known analytic expression for $M_{12}^s$.
As we use an effective Hamiltonian approach, our result for $\Gamma_{12}^s$ can
be applied to other NP models.
Our analysis involves a scan over a set of 10000 random points reproducing the correct low-energy spectrum
as well as the CKM mixing angles and phase. Bounds from $Z^0b\bar b$, $\eps_K$, and $\Delta m_{B_s}$ have been taken 
into account.  Due to the protection of the $Z^0 b_L\bar b_L$ vertex, the custodial extension allows for
bulk masses $c_{b_L}>-1/2$, which enlarges the contribution of RS penguin operators and $LL$ charged currents.  
While corrections to the magnitude and phase of $\Gamma_{12}^s$ turn out to be small, where for both RS variants
the biggest contribution comes from $\Q_2^{LR}$ for most of the allowed parameter space, the 
new CP violating phase in $M_{12}^s$ allows to relax the disagreement between theory and experiment.
Concerning the combined $\Delta\Gamma_s/\beta_s$ anaylsis, it is possible to enter the $68\%$ confidence
region. In order to reach the best-fit value however, moderate corrections to $|\Gamma_{12}^s|$ would be 
required \cite{Dighe:2010nj}, which are unlikely to appear in the models at hand.
For the case of the semileptonic CP asymmetry $A_{\rm SL}^s$, agreement can be obtained within $1\sigma$, where, 
at the same time, the best fit value of $S_{\psi\phi}$ can be reached.

\acknowledgments{We thank Martin Bauer, Sandro Casagrande, Uli Haisch, Tobias Hurth, Matthias Neubert, and Uli Nierste 
for useful discussions and remarks.}


\begin{appendix}

\section{Wilson coefficients of charged-current operators}
\label{app:charged}

The effective four-quark charged-current Hamiltonian can be written as
\begin{align}
&\mathcal{H}_\text{eff}^{(W)}=\,2\sqrt 2\, G_F \\
&\times\Big{\lbrace} 
[\,\bar d_{m_L}\gamma_\mu {({\bf V}_L^\dagger)}_{mn} u_{n_L} + 
\bar d_{m_R}\gamma_\mu {({\bf V}_R^\dagger)}_{mn} u_{n_R}]\nonumber\\
&\;\otimes
[\,\bar u_{m'_L}\gamma^\mu {({\bf V}_L)}_{m'n'} d_{n'_L} + 
\bar u_{m'_R}\gamma^\mu {({\bf V}_R)}_{m'n'} d_{n'_R}] (+ \text{h.c.})\Big{\rbrace}
\nonumber\,,
\end{align}
where $m,n,m',n'\in\lbrace1,2,3\rbrace$ and a summation over the repeated
indices is understood. Here, we have already absorbed a universal correction factor 
$(1+m_W^2/(2\Mkk^2)[1-1/(2L)])$ into the Fermi constant due to the 
normalization to muon decay, from which $G_F$ is extracted \cite{Casagrande:2008hr}. 
The tensor symbol merely indicates that the full analytic result contains terms that can not be
separated into independent matrix products. This is due to the sum over $W$-gauge boson 
profiles, which in the minimal model reads 
\begin{align}
\begin{split}
\displaystyle
2\pi\,& \sum_{n=0}\frac{\chi_n(t)\chi_n(t')}{m_n^2}=\frac 1{m_W^2}\\
&+\frac 1{2\Mkk^2}\left[L(t_<^2-t^2-t'^2)+1-\frac 1{2L}\right]\,,
\end{split}
\end{align}
where $t_<^2\equiv \text{min}(t^2,t'^2)\,$ \cite{Casagrande:2008hr}, and we dropped terms of
$\ord(v^4/\Mkk^4)$. The term $\propto t_<^2$ prevents a factorization into separate vertex factors.
Performing the overlap integrals with the corresponding fermion profiles
and employing the ZMA gives the rather simple result
\begin{align}\label{eq:LL}
\displaystyle
&{({\bf V}_L^\dagger)}_{mn}\otimes{({\bf V}_L)}_{m'n'} \\
&={(\bm{U}_d^\dagger\bm{U}_u)}_{mn}{(\bm{U}_u^\dagger\bm{U}_d)}_{m'n'}\left[1+\ord\left(\frac{v^2}{\Mkk^2}\right)\right] \nr \\
&+\frac{m_W^2}{2\Mkk^2}\,L\,{(\bm{U}_d^\dagger)}_{mi}{(\bm{U}_u)}_{in}
{(\bm{\widetilde\Delta}_{QQ})}_{ij} {(\bm{U}_u^\dagger)}_{m'j}{(\bm{U}_d)}_{jn'} \nr
\end{align}
with the non-factorizable correction \cite{Bauer:2008xb}
\beq\label{eq:Deltatilde}
\displaystyle
{(\bm{\widetilde\Delta}_{QQ})}_{ij}= \frac{F^2(c_{Q_i})}{3+2c_{Q_i}}\,\frac{3+c_{Q_i}+c_{Q_j}}
{2+c_{Q_i}+c_{Q_j}}\,\frac{ F^2(c_{Q_j})}{3+2c_{Q_ij}}\,.
\eeq
For $B_s^0$-meson decays, the whole expression has to be evaluated for $(m=2,n=2,m'=2,n'=3)$.
Here, the leading term in (\ref{eq:LL}), together with factorizable
corrections  of the form $v^2/\Mkk^2\, (...)_{mn}\cdot (...)_{m'n'}$ \cite{Casagrande:2010si}, 
should be identified with $\lambda^{bs}_c$. 
Concerning the custodial RS model, one would find additional 
factorizable terms, which also will be absorbed into CKM-matrix elements.
Thus, we find at LO in $v^2/\Mkk^2\,$
\begin{align}\label{eq:CLL}
\displaystyle
C_2^{LL}(\Mkk)&=\,\frac{m_W^2}{2\Mkk^2}\,L\\
&\times{\frac{{(\bm{U}_d^\dagger)}_{2i}{(\bm{U}_u)}_{i2}}{{(\bm{U}_d^\dagger\bm{U}_u)}_{22}}\,
{(\bm{\widetilde\Delta}_{QQ})}_{ij}\, 
\frac{{(\bm{U}_u^\dagger)}_{2j}{(\bm{U}_d)}_{j3}}{{(\bm{U}_u^\dagger\bm{U}_d)}_{23}} }\nr\,,
\end{align}
independent of the chosen scenario, and  $C_1^{LL}(\Mkk)=0\,$. The biggest corrections are found for $c_{Q_{2,3}}>-1/2$. 
For the mixed-chirality currents we have
\begin{align}
\displaystyle
\label{eq:LR}
&{({\bf V}_L^\dagger)}_{mn}\otimes\, {({\bf V}_R)}_{m'n'}\nr\\
&=\frac 1{\Mkk^2}{(\bm{U}_d^\dagger\bm{U}_u)}_{mn}
{(\bm{m}_u \bm{U}_u^\dagger)}_{m'j}\,
f(c_{Q_j})\, {(\bm{U}_d\,\bm{m}_d)}_{jn'}\,,\nr\\ \\
&{({\bf V}_R^\dagger)}_{mn}\otimes\, {({\bf V}_L)}_{m'n'}\nr\\ 
&=\frac 1{\Mkk^2} {(\bm{m}_d\bm{U}_d^\dagger)}_{mi}\,f(c_{Q_i})\,{(\bm{U}_u\,\bm{m}_u)}_{in}
{(\bm{U}_u^\dagger\bm{U}_d)}_{m'n'}\,.\nr
\end{align}
Here, $\bm{m}_u$ and $\bm{m}_d$ are $3\times 3$ diagonal matrices
containing the SM-like quark masses, and
\begin{align}
\begin{split}
f(c)=&\,\frac 1{F^2(c)(1-2c)}\,-\frac 1{1-2 c}\\
&\,+\frac{F^2(c)}{(1+2 c)^2}\left(\frac 1{1-2 c}-1+\frac 1{3+2 c}\right)\,.
\end{split}
\end{align}
Modifications due to the custodial model are of higher order.
We find the general RS prediction $C_1^{LR/RL}=0\,$ and
\begin{align}\label{eq:CLRCRL}
\begin{split}
& C_2^{LR}=\,\frac 1{\Mkk^2}\,
{\frac{{(\bm{m}_u \bm{U}_u^\dagger)}_{2i}\,f(c_{Q_i})\,{(\bm{U}_d\,\bm{m}_d)}_{i3}}
{{(\bm{U}_u^\dagger \bm{U}_d)}_{23}} }\,,\\
&C_2^{RL}=\,\frac 1{\Mkk^2}\,
{\frac{{(\bm{m}_d\bm{U}_d^\dagger)}_{2i}\,f(c_{Q_i})\,{(\bm{U}_u\,\bm{m}_u)}_{i2}}
{{(\bm{U}_d^\dagger\bm{U}_u)}_{22}} }\,,
\end{split}
\end{align}
where the coefficients should be matched at the KK scale.
The evolution down to the bottom mass is treated in Appendix \ref{sec:running}.

\section{Wilson coefficients of penguin operators}
\label{app:WilsonsP}

At $\ord(v^2/\Mkk^2)$ the Wilson coefficients of the penguin operators in 
equation (\ref{eq:Heff}) are explicitly given by
\cite{Bauer:2009cf} 
\begin{eqnarray}\label{eq:Cpenguin}
\begin{aligned}
   C_3^{\rm RS} &=
    \frac{\pi\alpha_s}{\Mkk^2}\,\frac{(\bm{\Delta}_D')_{23}}{2N_c} -
    \frac{\pi\alpha}{6\sws\cws\,\Mkk^2} (\,\bm{\Sigma}_D)_{23} \,,\\
%
   C_4^{\rm RS} &= C_6^{\rm RS} = -
    \frac{\pi\alpha_s}{2\Mkk^2}\,(\bm{\Delta}_D')_{23} \,, \\ 
%
   C_5^{\rm RS} &=
    \frac{\pi\alpha_s}{\Mkk^2}\,\frac{(\bm{\Delta}_D')_{23}}{2N_c} \,,\\
%
    C_7^{\rm RS} &= \frac{2\pi\alpha}{9\Mkk^2}\,(\bm{\Delta}_D')_{23} -
     \frac{2\pi\alpha}{3\cws\,\Mkk^2} (\,\bm{\Sigma}_D)_{23} \,, \\
%
    C_8^{\rm RS} &= C_{10}^{\rm RS} = 0
\,, \\
    C_9^{\rm RS} &= \frac{2\pi\alpha}{9\Mkk^2}\,(\bm{\Delta}_D')_{23} +
     \frac{2\pi\alpha}{3\sws\,\Mkk^2} (\,\bm{\Sigma}_D)_{23} \,, 
\end{aligned}
\end{eqnarray}
where
\beq\label{eq:Sigmas}
   \bm{\,\Sigma}_D \equiv  \omega_Z^{d_L} L \left( \frac12 - \frac{\sws}{3} \right)
   \bm{\Delta}_D + \frac{\Mkk^2}{m_Z^2}\,\bm{\delta}_D \,.
\eeq
These results are to be evaluated at the KK scale and are valid for the 
minimal RS variant for $\omega_Z^{d_L}=1$. 
In the custodial RS model with $P_{LR}$-symmetry, one finds 
$\omega_Z^{d_L}= 0\,$ \cite{Casagrande:2010si}.
Exact analytic expressions for $\bm{\Delta}_D$, $\bm{\Delta}_D'$, 
and $\bm{\delta}_D$ can be found in \cite{Casagrande:2008hr}.
However, as we only deal with light SM quarks in the initial and final 
state, it is convenient to apply the ZMA to the above expressions. Therefore
we have to replace
\beq
\begin{split}\label{eq:ZMA1}
   \bm{\Delta}_D &\to \bm{U}_d^\dagger\,\,\mbox{diag} \left[
    \frac{F^2(c_{Q_i})}{3+2c_{Q_i}} \right] \bm{U}_d \,, \\
   \bm{\Delta}'_D &\to \bm{U}_d^\dagger\,\,\mbox{diag} \left[
    \frac{5+2c_{Q_i}}{2(3+2c_{Q_i})^2}\,F^2(c_{Q_i}) \right]
    \bm{U}_d \,, 
\end{split}
\eeq
as well as
\begin{eqnarray}\label{eq:ZMA2}
\begin{aligned}
   &\bm{\delta}_D\to  \,\frac 1{\Mkk^2}\,\bm{m}_d\,\bm{W}_d^\dagger \\
   &\mbox{diag}\left[ \frac{1}{1-2c_{d_i}} 
    \left( \frac{1}{F^2(c_{d_i})} 
    - 1 + \frac{F^2(c_{d_i})}{3+2c_{d_i}} \right) \right] 
    \bm{W}_d\,\bm{m}_d \,. 
\end{aligned}
\end{eqnarray}
In the custodial model with extended $P_{LR}$-symmetry, the term $\propto 1/F^2(c_{d_i})$ 
in $\bm{\delta}_D$ is zero \cite{Casagrande:2010si}.
All other expressions hold for both scenarios.
The running of the penguin coefficients is also treated in Appendix \ref{sec:running}.

\section{Wilson coefficients for $\bm \Delta B=2$ operators}\label{sec:mixing}

The $\Delta B=2$ operators that contribute to the $\bar B_s^0$-$B_s^0$ mixing amplitude at 
tree-level are given by $\Q_1$, $\widetilde \Q_1$, $\Q_4$, and $\Q_5$. 
There is no mixing between $\Q_1$ and $\widetilde \Q_1$ under 
renormalization. The anomalous dimension for both cases is given by
$\gamma^{(0)\rm{VLL}}=6-6/N_c$ \cite{Buras:2000if}.
The opera\-tors $\Q_{4,5}$ mix under renormalization and
the anomalous dimension matrix can be taken from \cite{Bagger:1997gg,Buras:2000if}.
The running of the coefficients is described by the general formula (\ref{eq:rundown}).
Defining $\bm{\widetilde\Delta}_{dd}$ and $\bm{\widetilde\Delta}_{Qd}$ 
in analogy to (\ref{eq:Deltatilde}), the RS coefficients evaluated at the 
KK scale are given by \cite{Bauer:2009cf}
\begin{align}\label{eq:Cmix}
   C_1^{\rm RS} &=\, \frac{\pi L}{\Mkk^2}\, 
    {(\bm{U}_d^\dagger)}_{2i}{(\bm{U}_d)}_{i3}
    {(\bm{\widetilde\Delta}_{QQ})}_{ij} {(\bm{U}_d^\dagger)}_{2j}{(\bm{U}_d)}_{j3}\\
    \times&\left[
    \frac{\alpha_s}{2} \left( 1 - \frac{1}{N_c} \right) +
    Q_d^2\,\alpha + (\omega_Z^{d_Ld_L})\frac{(T_3^d-\sws\,Q_d)^2\,\alpha}
    {\sws\cws} \right] ,  \nr\\[2mm]
   \widetilde C_1^{\rm RS} &=\, \frac{\pi L}{\Mkk^2}\, 
    {(\bm{W}_d^\dagger)}_{2i}{(\bm{W}_d)}_{i3}
    {(\bm{\widetilde\Delta}_{dd})}_{ij} {(\bm{W}_d^\dagger)}_{2j}{(\bm{W}_d)}_{j3}\nr\\
    \times&\left[
    \frac{\alpha_s}{2} \left( 1 - \frac{1}{N_c} \right) +
    Q_d^2\,\alpha + (\omega_Z^{d_Rd_R})\frac{(\sws\,Q_d)^2\,\alpha}
    {\sws\cws} \right] , \nr\\[2mm]
   C_4^{\rm RS} &=- 2\alpha_s \frac{\pi L}{\Mkk^2}\, {(\bm{U}_d^\dagger)}_{2i}{(\bm{U}_d)}_{i3}
    {(\bm{\widetilde\Delta}_{Qd})}_{ij} {(\bm{W}_d^\dagger)}_{2j}{(\bm{W}_d)}_{j3}\nr\\[2mm]
   C_5^{\rm RS} &=\, \frac{\pi L}{\Mkk^2}\,
    {(\bm{U}_d^\dagger)}_{2i}{(\bm{U}_d)}_{i3}
    {(\bm{\widetilde\Delta}_{Qd})}_{ij} {(\bm{W}_d^\dagger)}_{2j}{(\bm{W}_d)}_{j3}\nr\\
    &\times\left[
    \frac{2\alpha_s}{N_c} 
    - 4 Q_d^2\,\alpha
    + \omega_Z^{d_Ld_R}
    \frac{4\sws\,Q_d\,(T_3^d-\sws\,Q_d)\,\alpha}{\sws\cws} \right] .\nr
\end{align}
Here we have introduced the correction factors $\omega_Z^{qq'}$, which are equal to 1
in the minimal RS model, and given by
\begin{align}
\begin{split}
\displaystyle
\omega_Z^{qq'} = 1+\frac 1{c_w^2-s_w^2}\,&\left(\frac{s_w^2(T_L^{3q}-Q^q)-c_w^2 T_R^{3q}}{T_L^{3q}-s_w^2 Q^q}\right)\\
		\times &\left(\frac{s_w^2(T_L^{3q'}-Q^{q'})-c_w^2 T_R^{3q'}}{T_L^{3q'}-s_w^2 Q^{q'}}\right)
\end{split}
\end{align}
in the custodial RS variant with $P_{LR}$-symmetry, where $s_w$ ($c_w$) is the sine 
(cosine) of the Weinberg angle. 
Numerically we find $\omega_Z^{d_Ld_L}\approx 2.9$, $\omega_Z^{d_Rd_R}\approx 150.9$, and $\omega_Z^{d_Ld_R}\approx -15.7 $.
The quantum numbers $T_R^{3q}$ can be found in \cite{Casagrande:2010si}, and $T_L^{3 d_L}\equiv T_3^d$.

\section{Running of the $\bm \Delta B=1$ coefficients}\label{sec:running}

Concerning the evolution of the RS Wilson coefficients, we will restrict ourselves to the
LO running in $\alpha_s$.
Within the operator basis $\vec Q=(Q_1,Q_2,Q_{3..10})$, the anomalous dimension matrix $\gamma^{(0)}$,
which is a function of $N_c$, $n_f$, $n_u$, and $n_d$ (number of colors, flavors, up- and down-type quarks),
can be found in \cite{Buras:1992tc,Ciuchini:1993vr}. 
The running of the coefficients is given by
\beq\label{eq:rundown}
\vec C(m_b)=U^{(5)}(m_b,m_t)\,U^{(6)}(m_t,\Mkk)\,\vec C(\Mkk)\,,
\eeq
where
\vspace{5mm}
\beq
\displaystyle
U^{(n_f)}(\mu_1,\mu_2)=
\hat V{\left({\left[\frac{\alpha_s^{(n_f)}(\mu_2)}
{\alpha_s^{(n_f)}(\mu_1)}\right]}^{\frac{\vec\gamma^{(0)}}{2\beta_0(n_f)}}\right)}_D \hat V^{-1}\,.
\eeq
Here, $\hat V$ diagonalizes ${\gamma^{(0)}}^T$ via
${\gamma^{(0)}}_D=\hat V^{-1} {\gamma^{(0)}}^T \hat V$,
and $\vec\gamma^{(0)}$ contains the entries of ${\gamma^{(0)}}_D$.
The QCD beta-function is given by $\beta_0(n_f)=(11 N_c-2\, n_f)/3$, and we
fix the running of $\alpha_s(\mu)$ at $\mu=m_t=171.2\,$GeV and $\mu=\Mkk=2\,$TeV.
As it turns out, there is a mixing between $Q_1$ and $Q_2$ independent of $n_f$, $n_u$, and $n_d$.
The evolution in the penguin sector gets a small admixture from charged currents.
The operators $Q_1^{LR/RL}$ and $Q_2^{LR/RL}$ do not mix into the penguin sector.
Their internal mixing is identical to that of the $LL$ operators, and there
is no mixing between charged currents of different chiralities. 
For the running of the $LR/RL$ coefficients, we insert
\beq
\displaystyle
\gamma^{(0)}=\left(
\begin{array}{cc}
  -\frac 6{N_c} & 6 \\
  6 & -\frac 6{N_c}
\end{array}\right)
\eeq
into equation (\ref{eq:rundown}), where this formula also holds for the $LL$ 
coefficients separately.

\end{appendix}

\end{document}